\documentclass[fleqn,usenatbib]{mnras}

\usepackage{newtxtext,newtxmath}

\usepackage[T1]{fontenc}

\DeclareRobustCommand{\VAN}[3]{#2}
\let\VANthebibliography\thebibliography
\def\thebibliography{\DeclareRobustCommand{\VAN}[3]{##3}\VANthebibliography}

\usepackage[dvipsnames]{xcolor}

\usepackage{graphicx}	
\usepackage{amsmath}	
\usepackage{hyperref}

\title[Sub-threshold joint GW-EM wave observation]{A simulation study on the sub-threshold joint gravitational wave-electromagnetic wave observation on binary neutron star mergers}

\author[Y. F. Du et al.]{
Y. F. Du,$^{1,2}$
Emre Seyit Yorgancioglu,$^{1,2}$
J. H. Rao,$^{1}$
Ankit Kumar,$^{1}$
S. X. Yi,$^{1}$\thanks{E-mail: sxyi@ihep.ac.cn }
S. N. Zhang,$^{1,2,3}$
and Shu Zhang$^{1}$
\\
$^{1}$Key Laboratory of Particle Astrophysics, Institute of High Energy Physics, Chinese Academy of Sciences, 19B Yuquan Road, Beijing 100049, People’s Republic of China\\
$^{2}$University of Chinese Academy of Sciences, Chinese Academy of Sciences, Beijing 100049, People’s Republic of China\\
$^{3}$National Astronomical Observatories, Chinese Academy of Sciences, Beijing 100012, People’s Republic of China
}

\date{Accepted XXX. Received YYY; in original form ZZZ}

\pubyear{\the\year{}}

\begin{document}
\label{firstpage}
\pagerange{\pageref{firstpage}--\pageref{lastpage}}
\maketitle

\begin{abstract}
The coalescence of binary neutron stars (BNS) is a prolific source of gravitational waves (GWs) and electromagnetic (EM) radiation, offering a dual observational window into the Universe. Lowering the signal-to-noise ratio (S/N) threshold is a simple and cost-effective way to enhance the detection probability of GWs from BNS mergers. In this study, we introduce a metric of the purity of joint GW and EM detections $P_{\rm joint}$, which is in analogue to $P_{\rm astro}$ in GW only observations. By simulating  BNS merger GWs jointly detected by the HLV network and EM counterparts (kilonovae and short Gamma-ray bursts, sGRBs) with an assumed merger rate density of BNS, we generate catalogs of GW events and EM counterparts. Through this simulation, we analyze joint detection pairs, both correct and misidentified. We find the following: 1. For kilonovae, requiring $P_{\rm joint}>$ 95\% instead of $P_{\rm astro}>95\%$ reduces the S/N from 9.2 to 8.5-8.8, allowing 5-13 additional joint detections per year and increasing the GW detection volume by 9-17\%;
2. For sGRBs, requiring $P_{\rm joint}>$ 95\% instead of $P_{\rm astro}$ reduces the S/N from 9.2 to 8.1-8.5;
3. Increasing kilonova or sGRB detection capability does not improve $P_{\rm joint}$ due to a higher rate of misidentifications.
We also show that sub-threshold GW and kilonova detections can reduce the uncertainty in measuring the Hubble constant to 89-92\% of its original value, and sub-threshold GW and sGRB observations can enhance the precision of constraining the speed of GWs to 88\% of previously established values.
\end{abstract}




\section{Introduction} \label{sec:intro}

Following the revelation of revolutionary discoveries through the observation of GW170817 and its electromagnetic counterpart (EMC) \citep[see][for a detailed review]{2017ApJ...848L..12A, 2021ARA&A..59..155M}, it is pertinent to acknowledge that, up to March 2024, the scientific community has yet to detect a second multi-messenger gravitational wave source (GW). This gap in discovery could likely be due to the present limitations of GW detectors.

GW sources in ground-based band accompanied by electromagnetic (EM) radiation are widely considered to originate from astrophysical phenomena such as mergers of binary neutron stars (BNS), black hole-neutron star (BH-NS) mergers, and binary black hole (BBH) mergers occurring within ambient environments ({\it e.g., } active galactic nucleus (AGN) disk) \citep[]{1998ApJ...507L..59L,2017ApJ...848L..12A, 2015ApJ...812...24F, 2017ApJ...835..165B, 2017MNRAS.464..946S,2019MNRAS.483.4197Y}. Consequently, augmenting the detection rate of BNS merger GW events is crucial for the expansion of multi-messenger source discoveries.
Currently, only two significant candidates for BNS merger GW have been observed, and the estimated rate of BNS merger based on GW observation is 10-1700 $\rm{Gpc^{-3}\,yr^{-1}}$ \citep{2023PhRvX..13a1048A}. There are two methods to increase the number of BNS merger GW detections: (1) improving the sensitivity of existing GW detectors or constructing new detectors with enhanced detection capabilities, and (2) lowering the GW detection threshold, though this approach may lead to a higher incidence of false signals. The most economical method under the current detection capabilities is the latter. The GW observation in the first half of the fourth observing run (O4a) has implemented sub-threshold observations. The alert threshold has been lowered to a FAR of 2 per day. As a result, the number of insignificant GW events is about 20 times that of significant GW events in LVK O4a public alerts. The characterization of insignificant GWs from compact binary sources have been discussed in \cite{2018PhRvD..98l3021H}. The precise degree to which the signal-to-noise ratio (S/N) threshold can be lowered is determined by false-alarm rate (FAR). 

In this work, we explore how electromagnetic counterparts (EMCs) can aid in identifying future gravitational wave events amid background noise. If the reduction of the threshold does not result in an excessive number of false multi-messenger sources during the follow-up of BNS merger GW events, we deem such sub-threshold GW observations to be acceptable. The purity of GWs, also written as $P_{\rm{astro}}$, serves as a metric for assessing the authenticity of a GW signal originating from an astrophysical source. Therefore, in this study, we introduce a novel definition for the purity of joint detection that characterizes the reliability of a multi-messenger signal from an astrophysical source when both GW and EM waves are detected. Through this revised definition of the purity, we are able to investigate the implications of reducing the S/N threshold in multi-messenger GW observations. Our idea is very similar to that of \cite{2024arXiv240708732V}, where a new $P_{\rm astro}$ is introduced with joint detection on sources with high-energy neutrinos and GWs.


Section 2 will describe the BNS model used for simulating BNS merger GWs and EMCs, along with the creation of the associated multi-messenger source catalogues. Section 3 will provide an overview of the background model for sub-threshold GW observations. Section 4 will introduce the newly defined purity and conduct simulations of EMC follow-up for GW candidates. Results from the simulations are set to be presented in Section 5, with Section 6 providing a thorough discussion of these findings.

\section{BNS population model and detect probabilities of multi-messenger counterparts} \label{sec:BNS}

The differential BNS event rate density as function of redshift $R(z)\equiv \frac{dN}{dtdz}$, can be written in terms of the volumetric total BNS merger rate $\mathcal{R}(z)\equiv \frac{dN}{dtdV_{c}}$ in the source frame as 

\begin{equation}
R(z)=\frac{1}{1+z}\frac{dV_{c}}{dz}\mathcal{R}(z),
\label{r_r}
\end{equation}
where $dV_{c}/dz$ is the differential comoving volume, and the $(1+z)^{-1}$ term arises from converting source-frame time to detector-frame time. $\mathcal{R}(z)$ is often refered to as the cosmic BNS merger rate density:
\begin{equation}
\mathcal{R}(z_m)=\mathcal{R}_n\int_{z_m}^{\infty}\psi(z_f)P(z_m|z_f)dz_f,
\label{merger_rate}
\end{equation}
where $\psi(z_f)$ is the Madau-Dickinson star formation rate:
\begin{equation}
\psi(z)=\frac{(1+z)^{\alpha}}{1+(\frac{1+z}{C})^{\beta}},
\label{Madau-Dickinson}
\end{equation}
with $\alpha=2.7$, $\beta=5.6$, $C=2.9$ \citep{2014ARA&A..52..415M}, and $P(z_m|z_f)$ is the probability that a BNS merges at $z_m$  given that the binary was formed at $z_f$.  This is referred to as the distribution of delay times, which takes the form from \cite{2019ApJ...886L...1V}: 
\begin{equation}
P(z_m|z_f,\tau)=\frac{1}{\tau}{\rm exp}[-\frac{t_f(z_f)-t_m(z_m)}{\tau}]\frac{dt}{dz}.
\label{prob_merger}
\end{equation}
In the above equation, $t_f$ and $t_m$ are the look back time as a function of $z_f$ and $z_m$, respectively. In this study, $\mathcal{R}_n$ and $\tau$ are the pair of parameters that define the merger rate density of the population. We selected the parameters for the BNS merger rate to be $\mathcal{R}_n=300\,\rm Gpc^{-3}\,yr^{-1}$ and $\tau=3\,\rm Gyrs$, consistent with the default values used in \cite{2022A&A...663A.155Y}. The $R(z)$ gives the redshift distribution of the BNS mergers. The definite integral of $R(z)$ can be used to determine the total number of BNS mergers within a specified redshift range. We set the upper limit of the redshift to 2.5, which is well beyond the detection range of BNS using LVK, kilonovae, and also larger than most of the detected redshift of sGRBs. For the distribution of Neutron Star (NS) masses, we use truncated Gaussian distributions with $m_{\rm{mean}}=1.4\,M_\odot$, $m_{\rm{scale}}=0.5\,M_\odot$, $m_{\rm{low}}=1.1\,M_\odot$, $m_{\rm{high}}=2.5\,M_\odot$, consistent with \cite{2022A&A...663A.155Y}. The right ascension, polarization angle, the cosine of declination, and the inclination angle are all assumed to be uniformly distributed within the Earth's equatorial coordinate system. Additionally, the spin of the neutron stars is not taken into account in this study. The arrival times of the binary neutron star mergers are also uniformly distributed across the observed time duration.

We consider the LIGO Handford, LIGO Livingston and Virgo (HLV) at their designed sensitivities as the GW observation network. Their sensitivity curves come from \href{https://dcc.ligo.org/LIGO-T1800044/public}{https://dcc.ligo.org/LIGO-T1800044/public}, see also \cite{2015CQGra..32k5012A} and \href{https://dcc.ligo.org/LIGO-T1500293/public}{https://dcc.ligo.org/LIGO-T1500293/public}. We use the triangulation method, where only the arrival times of the signals at all detectors are used to determine the network sky localization of the sources \citep{2011CQGra..28j5021F} (90\% probability region). Our simulations are limited to networks comprising at least three detectors. Therefore, most localization areas we simulate are smaller than those reported in LIGO/Virgo public alerts \citep{2022ApJ...924...54P}. Given the parameters of the BNS sources and the GW detector network, the S/N for a single detector can be calculated based on \cite{10.1093/acprof:oso/9780198570745.001.0001}, 
\begin{equation}
(\frac{S}{N})^2=4\int_{0}^{\infty}df\frac{{|\tilde{h}(f)|}^2}{S_{n}(f)},
\label{eq:snr}
\end{equation}
where $S_{n}$ denotes the one sided noise power spectral density of the detector.
The network S/N is calculated using $\sqrt{\sum_{i}\rho_{i}^2}$, 
where $\rho_{i}$ represents the S/N of a single detector. The BNS merger GWs are detected when the network's S/N exceeds the threshold. 

We assume that kilonovae and short Gamma-ray bursts (sGRBs) are exclusively generated by  BNS mergers. We focus on evaluating their detectability as a function of redshift. For kilonovae, the detection probability depends on factors such as the sensitivity of optical telescopes, observational strategies, and kilonovae templates \citep{2017ApJ...834...84C, 2022ApJ...927..163C, 2021MNRAS.504.1294S}. To simulate the detection probability as a function of redshift, we consider a simplified parametrized model. We suggest that the probability of detecting kilonovae can be represented by a simple piecewise function of redshift, similar to the detection efficiency of kilonovae described in \cite{2023ApJ...947...59L}
\begin{equation}
P_{\rm{det, kilo}}(z)=\left\{
\begin{array}{lr}
1 &  z\leq z_{\rm lim, kilo}; \\
{\rm exp}[-\frac{1}{2}(\frac{z-z_{\rm lim, kilo}}{\Delta z_{\rm kilo}})^2]  & z > z_{\rm lim, kilo}.\\
\end{array} \right.
\label{eq:Pdet_kilo}
\end{equation}
When $z\leq z_{\rm lim, kilo}$, kilonovae can be completely detected by the collaborative efforts of a number of optical telescopes. Beyond this limit, the probability of detection decreases exponentially. We select the 2.5 m Wide-Field Survey Telescope (WFST) and the Large Synoptic Survey Telescope (LSST) as the prototypes for the detection probability model parameters. Their exposure times are set at 40s and 200s, respectively. The efficiency of the i-band is chosen to represent the average detection capability. Based on the findings presented in Figure 8 of \cite{2023ApJ...947...59L}, for kilonovae resembling AT 2017gfo, the parameters governing their detection probability can be defined as $z_{\rm lim, kilo}=0.036$, $\Delta z_{\rm kilo}=0.022$, and $z_{\rm lim, kilo}=0.067$, $\Delta z_{\rm kilo}=0.056$, respectively. 

For sGRBs, we use a similar method as in \cite{2020ApJ...894...11G} to get the detection probability. We assume a structured Gaussian-type jet model. Then the luminosity per solid angle in the observer frame is
\begin{equation}
    L(\theta_v)=\frac{dE}{dtd\Omega}=\frac{L_0}{4\pi}e^{-\theta_v^2/\theta_c^2},
    \label{eq:Lum}
\end{equation}
where the $\theta_c$ is the jet’s core angle, $L_0$ is the isotropic-equivalent luminosity observed on-axis, and $\theta_v$ is viewing angle \citep{2019MNRAS.485.1435H}. The flux for a GRB at redshift $z$ and viewing angle $\theta_v$ is:
\begin{equation}
F(\theta_v,z)=\frac{L(\theta_v)}{4\pi D^2_{L}(z)},
\end{equation}
where $D_{L}(z)$ is the luminosity distance converted from $z$ by a cosmological model found in \cite{2020A&A...641A...6P}. Now we assume that the detection flux threshold is $F_{\rm lim}$. For \textit{Fermi} Gamma-ray Burst Monitor (GBM),  we take its 64 ms limiting flux, $F_{\rm lim}=2.0\times 10^{-7}$\,erg\,cm $^{-2}$\,s$^{-1}$ \citep{2017ApJ...848L..14G}. Then the minimum $L(\theta_v)$ can be detected is
\begin{equation}
L_{\rm lim}(\theta_v)=F_{\rm lim}4\pi D_{L}^2(z),
\end{equation}
that means $L_0$ should satisfy
\begin{equation}
L_0 \ge L_{\rm{lim}}\equiv F_{\rm{lim}}(4\pi)^2 D_{L}^2(z)\exp(\theta^2_v/\theta_c^2).
\end{equation}
The $L_0$ distribution of the structured jet in the sGRB has been investigated in a study by \cite{2023A&A...680A..45S}. Based on their research, we adopt a power law with an index of $-A$, with a lower exponential cutoff below $L_{\star}$ in the distribution of $L_0$ as follows
\begin{equation}
\Phi(L_0)=\frac{\Phi_0 A}{L_{\star}(1-1/A)}\exp\left[-\left(\frac{L_{\star}}{L_0}\right)^A \right]\left(\frac{L_0}{L_{\star}}\right)^{-A},
\label{dis_L}
\end{equation}
where $\Phi_0$ is the normalization parameter. Following the paper of \cite{2023A&A...680A..45S}, we take the values $A=3$ and $L_{\star}=0.5\times 10^{52}$\,erg\,s$^{-1}$.
Then the detection probability is:
\begin{equation}
\label{eq:p_det1}
P_{\rm{det, sGRB}}(\theta_v,z)\propto \frac{\int_{L_{\rm{lim}}}^\infty\Phi(L_0)dL_0}{\int_{0}^\infty\Phi(L_0)dL_0},
\end{equation}
In equation \ref{eq:p_det1}, the results of the two definite integrals are the gamma function $\Gamma$ and the incomplete gamma function $\widetilde{\Gamma}$, respectively. Thus, we get
\begin{equation}
P_{\rm{det, sGRB}}(\theta_v, z)=\Lambda \frac{\widetilde{\Gamma}(1-1/A, (L_{\star}/L_{\rm lim})^A)}{\Gamma(1-1/A)},
\label{eq:Pdet_grb}
\end{equation}
where $\Lambda=0.6$, which means a 60 percent sky coverage (time-averaged) for \textit{Fermi} GBM telescope following \cite{2016ApJ...818..110B}. We set $\theta_c=1.2^{\circ}$, which is in accordance with the observational implied range from 1$^{\circ}$ to 9$^{\circ}$ \citep{2018Natur.561..355M}. The resultant {\it Fermi} GBM detection rate and redshift range are in good agreement with reality (40 sGRB/year and up to $z\sim2.5$). We consider another GRB detector with better capability, with flux limit $F_{\rm lim}=2.0\times 10^{-8}$\,erg\,cm $^{-2}$\,s$^{-1}$ and whole sky coverage, corresponding to Gravitational wave high-energy Electromagnetic Counterpart All-sky Monitor (GECAM) in its designed capability \citep[]{2022MNRAS.511..964X}. Therefore, we can get the detection probability of each BNS merger sGRB by equation \ref{eq:Pdet_grb}.

\begin{figure*}
    \centering
    \includegraphics[width=\textwidth]{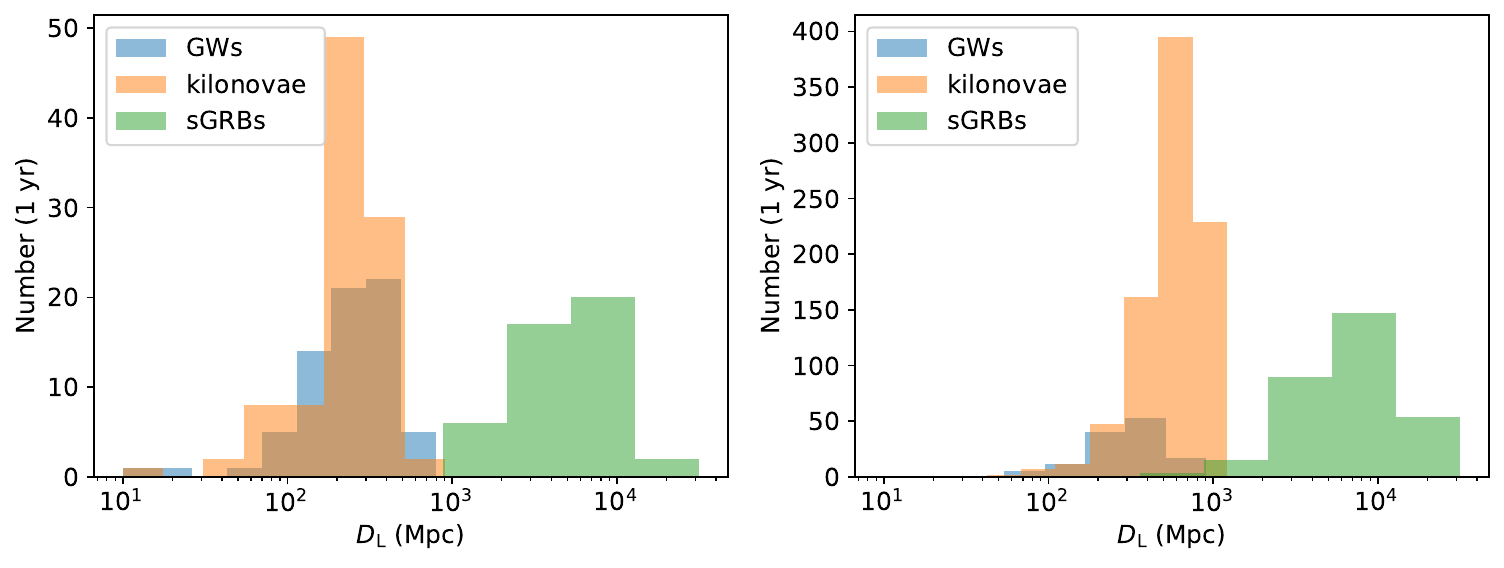}
    \caption{The luminosity distance histogram of the GWs, kilonovae and sGRBs observed over one year. The HLV network's S/N threshold for the left and right panels are 10 and 8, respectively. On the left panel, the kilonova detector is WFST; while on the right panel, it is LSST. The GRB detector for the left panel is \textit{Fermi} GBM and GECAM for the right panel.}
    \label{fig:BNS}
\end{figure*}

Figure \ref{fig:BNS} presents the luminosity distance distribution for GWs, kilonovae, and sGRBs originating from the population of BNS mergers. From their  $z$-distribution, we can infer that the detected kilonovae and GW BNS events largely overlap. In contrast, only a small fraction of sGRBs are detectable as EMCs of GWs, aligning with the findings reported in \cite{2023A&A...672A..74H}.

\section{GW background model}
\label{sec:back}

If we lower the S/N threshold, $\rho_\star$, more noise from the detector's background will be misidentified as signals. Consequently, the FAR increases with a lower $\rho_\star$. \cite{2018ApJ...861L..24L} found that the FAR for the LIGO Hanford and LIGO Livingston detectors during the O1-O2 observing runs, as a function of $\rho_\star$, can be accurately modeled using an exponential function:
\begin{equation}
\rm{FAR}(\rho_{\star})=\rm{FAR}_8\times\exp\big[-\frac{\rho_\star-8}{\alpha}\big],
\label{LynchFAR}
\end{equation}
where FAR$_8$ is the FAR when $\rho_\star=8$, and $\alpha$ is the parameter used to adjust the distribution of background sources. For BNS, \cite{2018ApJ...861L..24L} found $\alpha$=0.13 in equation \ref{LynchFAR}, arguing that this value is representative of BNS LIGO Livingston and LIGO Hnford (HL) searches across all runs. They found that the FAR$_8$ for BNS from O1-O2 is $3\times10^4$\,yr$^{-1}$, which we take as a reference value for our simulation, although it may vary from run to run. When a third detector is included in the network, $\alpha$ will become smaller. It is equivalent that for three detector network, the imaginary BNS population has less overall merger rate and steeper $z$ dependence \citep{2018ApJ...861L..24L}. In real GW runs, however, a trigger has been identified based on HL coincidence detection criteria, even in a three-detector run. Therefore, although we are simulating a HLV network, we will still use the background population model in the following way: when simulating the detection of fake signals, we draw samples from the imaginary BNS population with parameters corresponding to HL search as found in \cite{2018ApJ...861L..24L}. The network SNR of fake source does not include the contribution from Virgo, while the localization of the source is still based on three detectors.

\begin{figure*}
    \centering
    \includegraphics[width=.8\textwidth]{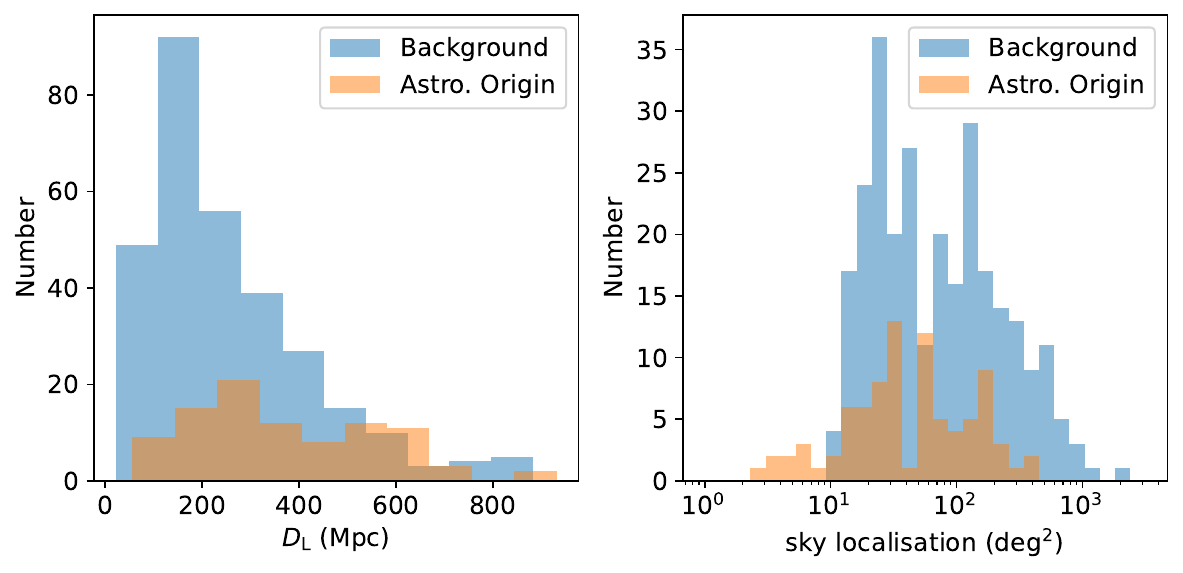}
    \caption{Histograms of luminosity distance and sky localization area for background GW sources and astrophysical origin GW sources over one year. The S/N thresholds are both 8.}
    \label{fig:are}
\end{figure*}

Equation \ref{LynchFAR} provides the number of background GW sources detected by the HLV network in one year. We further assume that this BNS population, mimicked by noise, has a uniform chirp mass distribution within the search window. According to the formula (7.183) on of \cite{10.1093/acprof:oso/9780198570745.001.0001}, we have
\begin{equation}
\rho = A\frac{\mathcal{M}_c^{5/6}}{D_{\rm L}},
\label{rho_MD}
\end{equation}
where $D_{\rm L}$ is the luminosity distance of the source, $\rho$ is the S/N of the source, $A$ is a constant if the detectors are identical, and $\mathcal{M}_{c}$ (where $\mathcal{M}_{c}=(m_1 m_2)^{3/5}(m_1 + m_2)^{-1/5}$) is the chirp mass. Using this equation, we can quickly calculate the $D_{\rm L}$ from $\rho$, $\mathcal{M}_{c}$ and $A$. To simulate the sources of the GW background, we took the following steps:
First, we assume the distribution of $\mathcal{M}_c$ for GW sources is uniform from 0.435 to 2.61 solar masses (M$_\odot$), corresponding to the NS mass range of 0.5-3 M$_\odot$, which represents a conservative range for NS mass \citep{2013ApJ...778...66K} in waveform searches. To simplify the calculations, we assumed equal masses for the two NS, thereby simultaneously determining the masses of the NS.
Second, we randomly assign the right ascension, declination, polarization angle, and the inclination angle to the background sources in the same manner as we generate the BNS merger population. The arrival times of the sources are uniformly distributed over one year. 
Third, we assume that the $\rho$ for all background sources is $\rho_{\star}$. 
Then, the $D_{\rm L}$ can be inferred from their $\rho$, $\mathcal{M}_c$, and angles using Equation \ref{eq:snr} and \ref{rho_MD}.
Finally, We apply the method described in Section \ref{sec:BNS} to determine the HLV network sky localization of the sources and use the approximation $\Delta D \sim D/\rho$ to estimate the error in $D_{\rm L}$ \citep[]{1994PhRvD..49.2658C}.
In Figure \ref{fig:are}, we plot histograms of the luminosity distance and sky localization for both BNS merger GW sources and GW background sources. It is evident that the sky localization of the background GWs is broader than that of GWs of astrophysical origin, and there is little difference in the luminosity distance distribution between these two types of sources.

\section{Simulation of multi-messenger GW observations}

\begin{figure}
    \centering    \includegraphics[width=.45\textwidth]{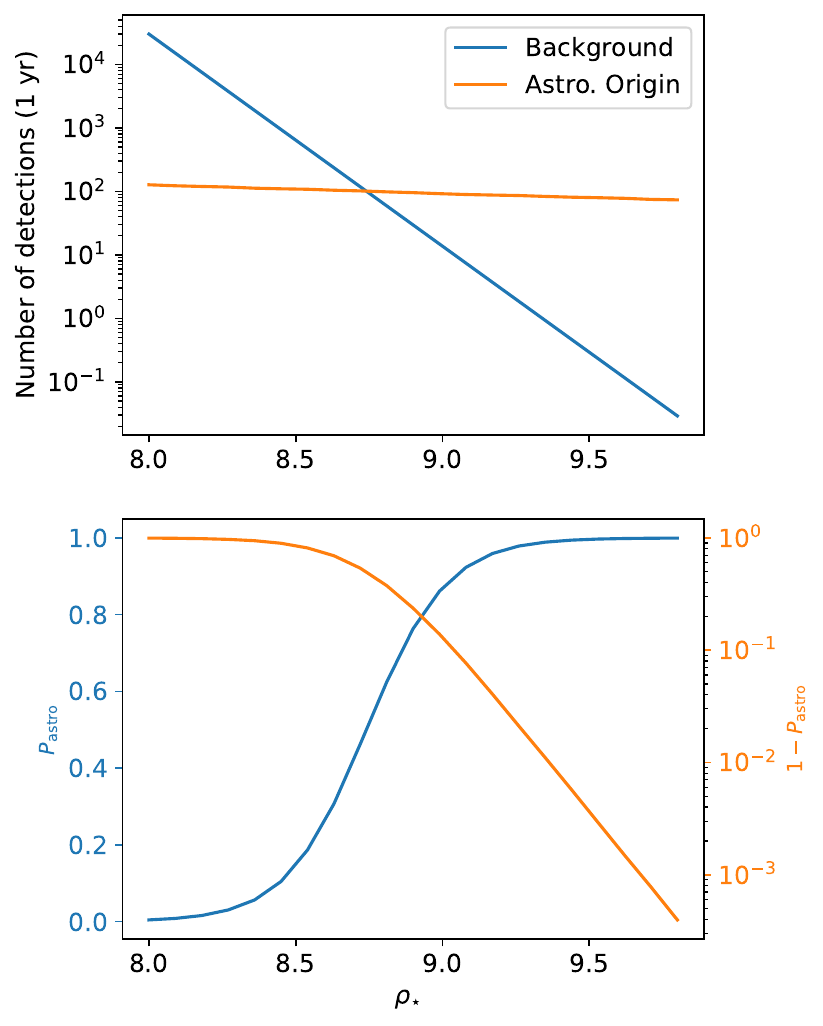}
    \caption{The astrophysical origin GW and background GW as functions of the network S/N threshold. The $P_{\rm astro}$ and $1- P_{\rm astro}$ as functions of the network S/N threshold.}
    \label{fig:2figs}
\end{figure}

From Sections \ref{sec:BNS} and \ref{sec:back}, we can generate catalogues for GW signals of both astrophysical origin and background, as well as for kilonovae and sGRBs. The purity of the GW signals is defined as
\begin{equation}
P_{\rm{astro}} (\rho_{\star})=\frac{N_{\rm{GW, astro}}(\rho_{\star})}{N_{\rm{GW, astro}}(\rho_{\star})+N_{\rm{GW, back}}(\rho_{\star})},
\label{astro}
\end{equation}
where $N_{\rm{GW, astro}}$ and $N_{\rm{GW, back}}$ represent the number of astrophysical origin GW signals and background GW signals, respectively. We plot $N_{\rm{GW, astro}}$, $N_{\rm{GW, back}}$ over one year, and $P_{\rm{astro}}$ as functions of $\rho_{\star}$ in Figure \ref{fig:2figs}. The figure illustrates that the number of background sources is almost equal to that of astrophysical sources when $\rho_{\star}$ is approximately 8.8. To ensure that the number of background sources is at least one and to avoid an excessive number of background sources that would significantly increase the simulation time, we select values of $\rho_{\star}$ from 8 to 9.3 at intervals of 0.1 for the simulation.

Our understanding of the observations of GWs and EMCs from BNS mergers has been greatly enriched by insights gained from the first multi-messenger observations of a BNS merger \citep{2017ApJ...848L..12A}. Typically, GW signals are detected first, followed within a few seconds by the detection of an sGRB by high-energy detectors. Subsequently, the optical counterpart (kilonova or optical afterglow) becomes observable, ranging from hours to weeks after the initial GW signal. In light of this sequence, we use the arrival time, luminosity distance, luminosity distance error, and sky localization of the GWs to identify the s also have corresponding EMCs. During EM follow-up observations, we also need to take into account the sky localization uncertainties of sGRBs, ranging from a few square degrees to several hundred square degrees \citep{2015ApJS..216...32C}. In this study, we assume a sky localization uncertainty of 75 square degrees (68\% confidence level) for each sGRB. If the localization of a GW event overlaps with the localization of a sGRB, we classify that these detections in spatial coincidence.
For the sake of simplicity, this analysis presumes that the localization region forms a perfect circle, despite the fact that it typically resembles an ellipse in actuality. 
For a kilonova or sGRB, we consider it as candidate of EMC of a GW if its arrival time and sky localization is within the spacetime coincidence window of the GW signal.

In our work, we assume that each BNS merger GW, including the sub-threshold GW signals, will be followed up. We thus perform the following steps:
\begin{enumerate}

\item We use a BNS population model and detection probabilities of multi-messenger counterparts to produce the BNS merger GW, kilonova, and sGRB catalogues, and employ a GW background model to generate the GW background catalogues for different $\rho_{\star}$ values over one year.
\item For each GW in the BNS merger GW catalogues and GW background catalogues for a specific $\rho_{\star}$, we record their luminosity distance $D_{{\rm L}, i}$, its error $\Delta D_{{\rm L}, i}$, sky localization area $\Delta \Omega_i$, sky coordinates ($\theta_i$, $\varphi_i$), and detection time $T_i$. For each EMC in the BNS merger kilonova catalogues or sGRB catalogues for a specific $\rho_{\star}$, we note their luminosity distance $D_{{\rm L}, j}$, sky coordinates ($\theta_j$, $\varphi_j$), sky localization area $\Delta \Omega_j$, and detection time $T_j$.
\item For the $i$-th GW, we identify a pool of potential EMCs $\{E_j\}_i$ from the overall kilonova or sGRB catalogue. The EMC pool consists of EMCs for which $D_{{\rm L}, i}-\Delta D_{{\rm L}, i}< D_{{\rm L}, j}< D_{{\rm L}, i}+\Delta D_{{\rm L}, i}$ (this condition does not need to be met for sGRBs because the distance measurement for sGRBs is very inaccurate in most cases), $\delta\Omega_{ij}<\sqrt{\Delta\Omega_i/\pi}+\sqrt{\Delta\Omega_j/\pi}$ (this condition represents the sky localization coincidence, where $\sqrt{\Delta\Omega_j/\pi}$ approximates the radius of localization circle of the $i$-th source), and $\delta t_{ij}$ within $\Delta T_{\rm EMC}$, where $\delta\Omega_{ij}$ is the angular separation between the $i$-th GW candidate and the $j$-th EMC, $\delta t_{ij}$ is the time lag between the $j$-th EMC and the $i$-th GW candidate, and $\Delta T_{\rm EMC}$ is the coincident time window between a GW and its potential EMC. The $\Delta T_{\rm EMC}$ is -2 to 10 seconds for sGRBs and 0 to one week for kilonovae, respectively.
\item From this pool, we randomly assign one EMC to the $i$-th GW.
 
\end{enumerate}

\begin{figure}
    \centering
    \includegraphics[width=.45\textwidth]{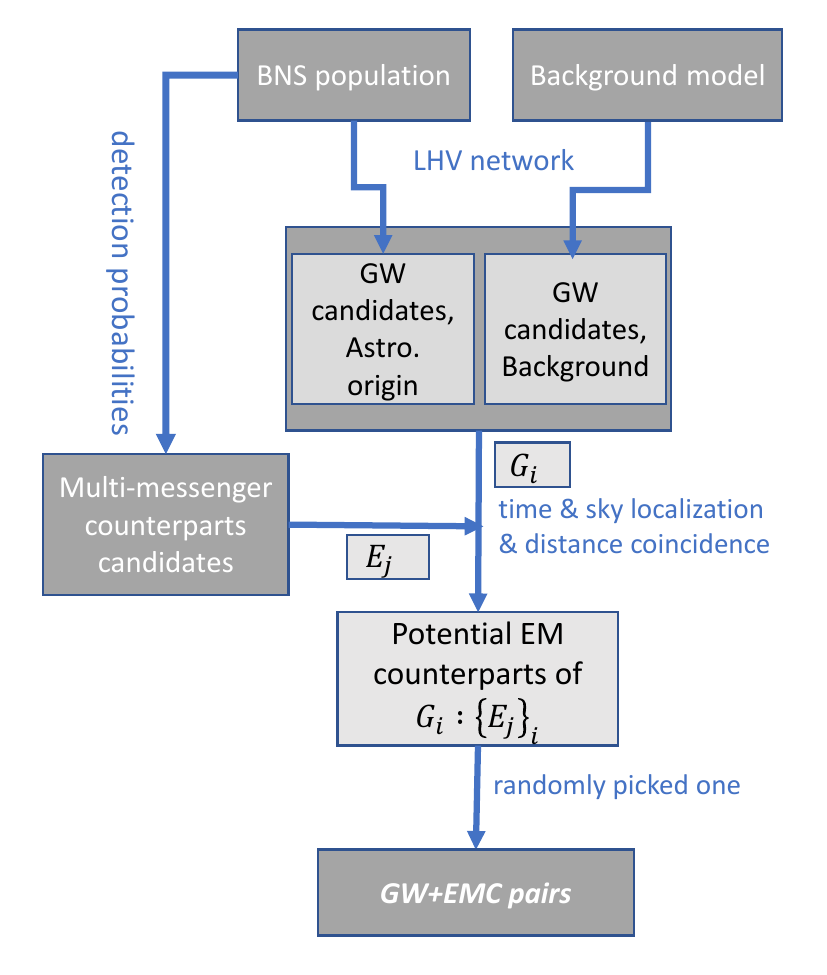}
    \caption{The sketch of simulation multi-messenger GW observation.}
    \label{fig:sketch}
\end{figure}

We summarize these steps in Figure \ref{fig:sketch}.
Pairing GWs with EMCs can lead to misidentification, which falls into three categories: 1.\ Even if the space-time coincidence of GW and EMC is met, the two signals may originate from separate astrophysical events. 2.\ The identified GWs are actually background sources rather than genuine signals; 3.\ The EMCs are contaminants, which include those not originating from BNS mergers. In this work, we have not simulated background EMC sources. Therefore, we only consider the first two scenarios. After the above steps, we simulated numerous GW and EMC pairs obtained from multi-messenger detection. For these pairs, we define a new purity:
\begin{equation}
P_{\rm joint}(\rho_{\star})=\frac{N_{\rm{GW+EMC}}^{\rm{astro}}(\rho_{\star})}{N_{\rm{GW+EMC}}^{\rm{astro}}(\rho_{\star})+N_{\rm{GW+EMC}}^{\rm{misid}}(\rho_{\star})},
\label{astro_new}
\end{equation}
Where $N_{\rm{GW+EMC}}^{\rm{astro}}$ represents the number of genuine pairing pairs, and $N_{\rm{GW+EMC}}^{\rm{misid}}$ represents the number of misidentified pairs.
\section{Results}

\begin{figure}
    \centering
    \includegraphics[width=.45\textwidth]{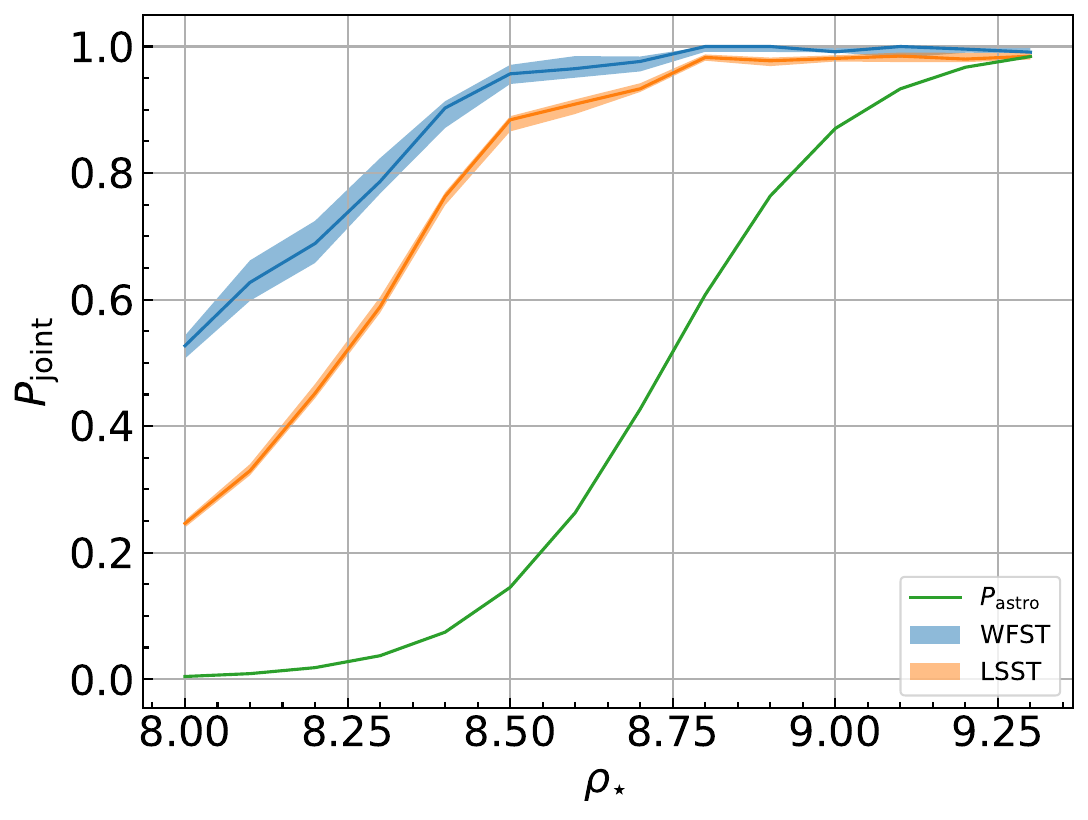}
    \caption{The $P_{\rm joint}$ versus $\rho_{\star}$ for two different sets of $P_{\rm{det, kilo}}$ parameters. The blue area represents WFST, while the orange area represents LSST. The shaded area represents the 90\% confidence interval, and the solid lines within this area are the median lines. The green curve represents $P_{\rm{astro}}$ calculated using equation \ref{astro}.}
    \label{fig:pastronew}
\end{figure}

\begin{figure*}
    \centering
    \includegraphics[width=\textwidth]{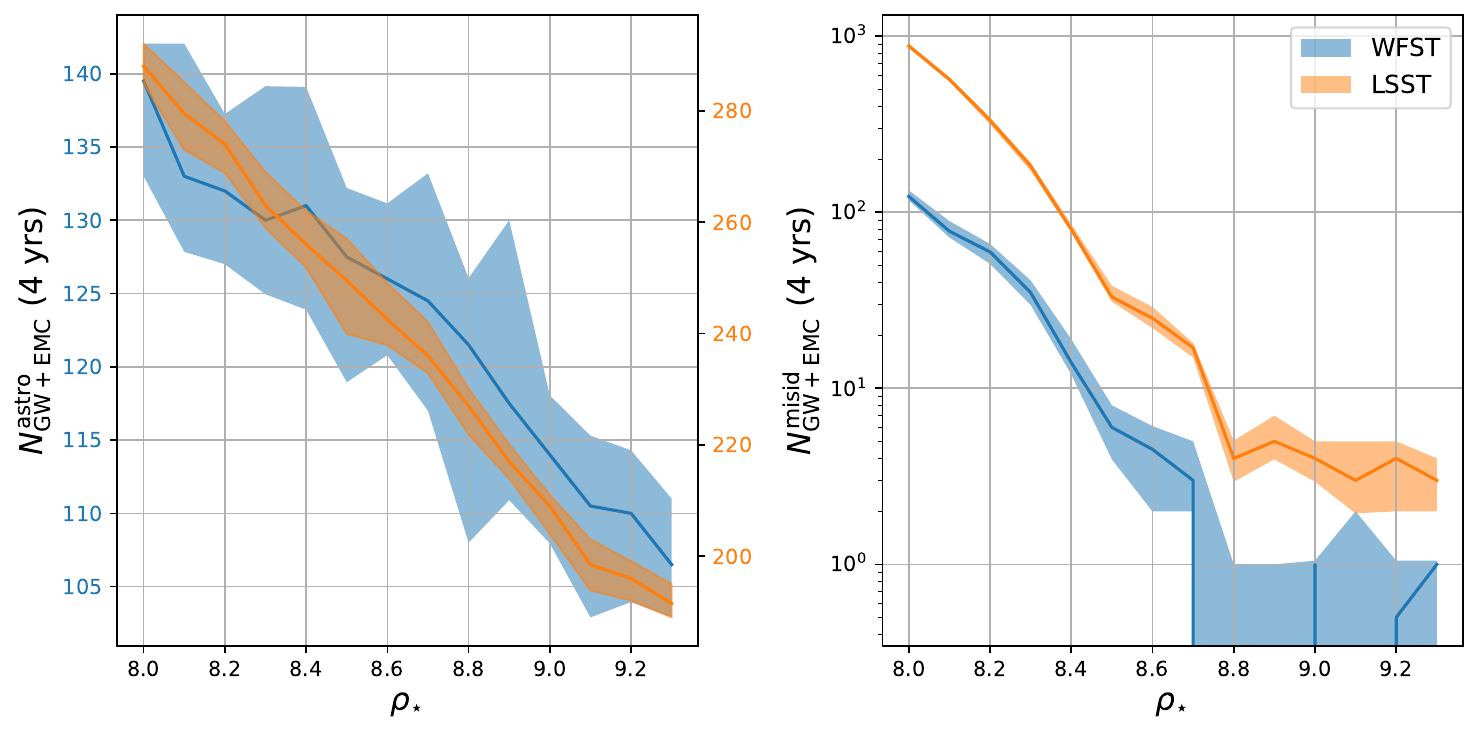}
    \caption{The left panel shows the number of genuine joint GW and kilonova multi-messenger sources versus $\rho_{\star}$, and the right panel shows the number of misidentifications in GW and kilonova pairing simulations versus $\rho_{\star}$ for two different sets of $P_{\rm{det, kilo}}$ parameters. One set corresponds to WFST, while the other corresponds to LSST. The detection duration is 4 years. The shaded area represents the 90\% confidence interval, and the solid lines are the median lines. In the left panel, we use two different y-axes to facilitate a clear comparison between the two sets of data.}
    \label{fig:num}
\end{figure*}

We make many realizations of simulated catalogues with the same setting for each S/N threshold to average out the uncertainty due to the randomness in different realisations. In this study, we use 20 realizations, which we find sufficient to provide robust upper and lower limits for $P_{\rm joint}$ while keeping computational costs manageable. The observation duration is 4 years. Figure \ref{fig:pastronew} presents the $P_{\rm joint}$ results obtained using kilonova detection parameters from two different sets of simulations. Figure \ref{fig:num} presents the $N_{\rm{GW+EMC}}^{\rm{astro}}$ and $N_{\rm{GW+EMC}}^{\rm{misid}}$ results obtained using kilonova detection parameters from two different sets of simulations. The zig-zag behavior in Figure \ref{fig:num} is due to the randomness in the simulation samples. While further increasing the sample size could reduce this randomness, we did not do so to limit computational costs. From Figure \ref{fig:pastronew}, we observe that if the $P_{\rm joint}$ threshold is set to 95\%, we can reduce $\rho_{\star}$ from 9.2 to 8.5 or 8.8. If the $P_{\rm joint}$ threshold is set to 60\%, we can reduce $\rho_{\star}$ from 8.8 to 8.1 or 8.3. At the same time, we find that improvements in future kilonova detection capabilities do not allow us to lower $\rho_{\star}$ further; instead, they require a higher $\rho_{\star}$. This issue is explained in Figure \ref{fig:num}. It can be observed that increasing kilonova detection capabilities introduces a higher proportion of misidentifications compared to correct multi-messenger sources. 
Furthermore, Figure \ref{fig:num} also shows that if $\rho_{\star}$ is from 9.2 reduced to 8.5, the increase in the number of multi-messenger sources in 4 years for the two sets of parameters is approximately 20 and 52, respectively, while the number of misidentified sources in 4 years increases from 1 and 3 to approximately 6 and 30, respectively. Additionally, as $\rho_{\star}$ decreases, misidentifications grow exponentially similar to background sources, while the increase in multi-messenger sources tends towards linearity, with a tendency to flatten. Therefore, $\rho_{\star}$ can only be lowered to between 8.5 and 8.8 for the GWs and kilonovae multi-messenger observation simulation.

\begin{figure}
    \centering
    \includegraphics[width=.45\textwidth]{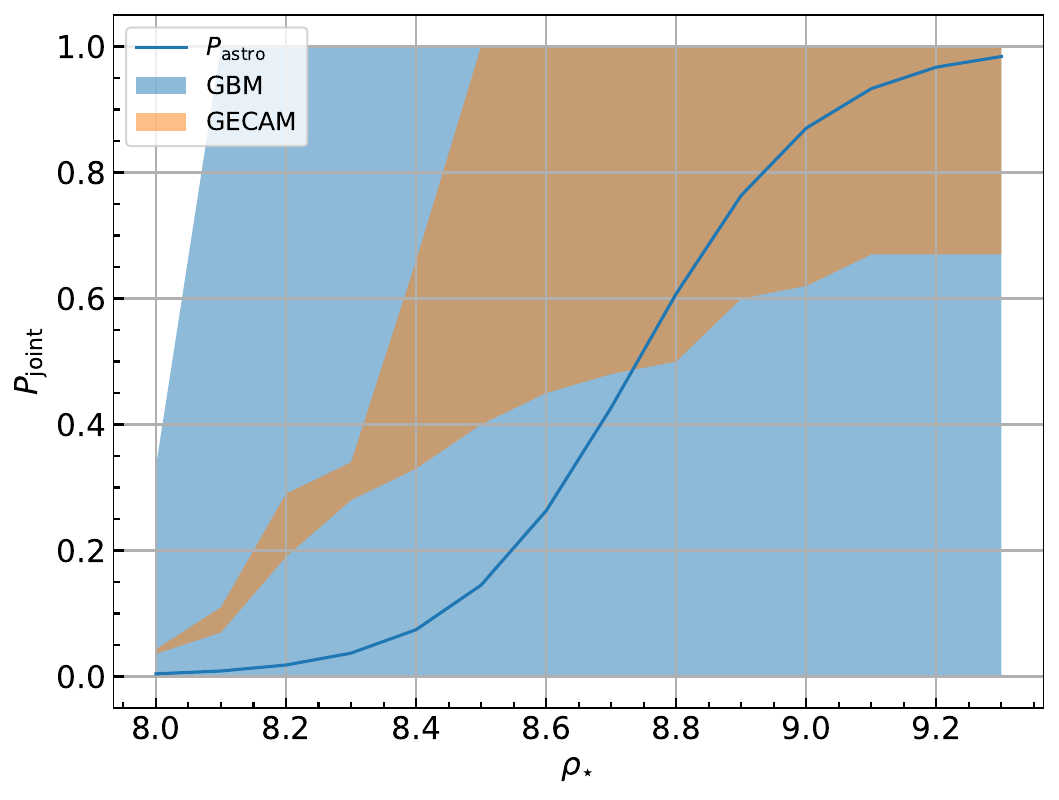}
    \caption{The $P_{\rm joint}$ versus $\rho_{\star}$ for two different sets of $P_{\rm{det, kilo}}$ parameters. The blue area represents \textit{Fermi} GBM, while the orange area represents GECAM. The shaded area represents the 90\% confidence interval. The blue curve represents $P_{\rm{astro}}$ calculated using equation \ref{astro}.}
    \label{fig:pastronew_sg}
\end{figure}

We find that the joint GW and sGRB source rate ranges from 0 to 0.25 yr$^{-1}$ and 0.5 yr$^{-1}$ for \textit{Fermi} GBM and GECAM, respectively.
Simultaneously, we find that lowering the S/N threshold from 9.2 to 8 increased the joint detection rate by approximately 0.1 yr$^{-1}$ for \textit{Fermi} GBM.
Despite using sub-threshold GW and sGRB observations, the joint GW and sGRB detection rate is too low to show a significant increase.
Figure \ref{fig:pastronew_sg} presents the $P_{\rm joint}$ results under two gamma-ray detectors. From Figure \ref{fig:pastronew_sg}, we observe a particularly large confidence interval for $P_{\rm joint}$. This is due to the low detection probability, resulting in many realizations where the genuine joint detection number is zero for which we can only estimate a lower limit for $P_{\rm joint}$. Additionally, we find the same pattern in Figure \ref{fig:pastronew_sg} as in Figure \ref{fig:pastronew}, where instruments with higher sensitivity do not exhibit smaller $P_{\rm joint}$ values at the same $\rho_{\star}$. The reason is analogous to the joint detection of GW and kilonova, where the rate of misidentifications in joint detections grows exponentially. If the $P_{\rm joint}$ threshold is set to 95\%, the results from Figure \ref{fig:pastronew_sg} indicate that for \textit{Fermi} GBM, $\rho_{\star}$ should be greater than 8.1, and for GECAM, it should be greater than 8.5.


We conclude as follows:
\begin{itemize}
    \item If the purity threshold is set to 95\%, changing from $P_{\rm{astro}}$ to $P_{\rm{joint}}$ can reduce the $\rho_{\star}$ from 9.2 to 8.5 or 8.8 for kilonovae as the EMCs of the GWs. If the purity threshold is set to 60\%, $\rho_{\star}$ can be reduced from 8.8 to 8.1 or 8.3 for kilonovae as the EMCs of the GWs. The two scenarios represent WFST and LSST, respectively.
    
    \item We find that, compared to WFST, LSST does not have a larger $P_{\rm joint}$ at the same $\rho_{\star}$. On the contrary, due to the introduction of more kilonovae contamination, the $P_{\rm joint}$ will be smaller.

    \item If we lower the $\rho_{\star}$ from 9.2 to 8.5 for joint GW and kilonova observations under two optical telescopes, we can increase the number of astrophysical and correctly identified joint GW and kilonova observations from 27 or 49 to 32 or 62 per year. The increase number of joint multi-messenger observations is 5 or 13 per year, respectively. 
    
    \item If we choose \textit{Fermi} GBM and GECAM as the GRB detectors, the joint observation rate is 0-0.25 yr$^{-1}$ and 0.5 yr$^{-1}$, respectively. If the purity threshold is set to 95\%, changing from $P_{\rm{astro}}$ to $P_{\rm{joint}}$ can reduce the $\rho_{\star}$ to 8.1 for GBM and 8.5 for GECAM in optimal cases.

\end{itemize}
\section{Discussion}

Bright sirens can be used to constrain the Hubble constant $H_0$ when we have the luminosity distance from GWs and redshift measurement from their optical counterparts \citep[][]{1986Natur.323..310S, 2017Natur.551...85A}. In this work, we find that the number of bright sirens in one year can increase from 27 or 49 to 32 or 62 if we lower the $\rho_{\star}$ from 9.2 to 8.5 for WFST and LSST, respectively. According to \cite{2018Natur.562..545C}, the fractional $H_0$ measurement uncertainties, defined as half the width of the symmetric 68\% credible interval divided by the median, are proportional to $1/\sqrt{N}$, where $N$ is the number of bright sirens. Therefore, the fractional $H_0$ measurement uncertainties could be reduced to 92\% or 89\% of the original uncertainties if we utilize sub-threshold GWs and kilonovae joint observations. The above estimation is rough, and did not consider that $\Delta D_{\rm L}/D_{\rm L}$ of the sub-threshold sources can be significantly larger.


On the other hand, we can use the arrival time delay between one GW and its sGRB counterpart to constrain the velocity of GW and Lorentz violation \citep[]{2017PhRvL.119p1102C, 2016PhLB..757..510K, 2024arXiv240513314R}. The most accurate limit depends on the farthest GW and sGRB joint detection event. Lowering the S/N threshold of GW can extend the observation distance of GW based on equation \ref{rho_MD}. Therefore, if we lower the S/N threshold from 9.2 to 8.1, the GW observation distance limit will increase by a factor 1.14. The accuracy of limiting the velocity of gravitational waves will be improved to 88\% of former \citep{2024arXiv240513314R}. 

Based on the findings of this study, we recommend that the S/N threshold for GWs be lowered to 8.5-8.8 (corresponding to a FAR of 63-641 per year) for kilonova and GW joint observations. For these low-threshold BNS merger GW signals, certain resources can be allocated for follow-up observations. The increase in the number of multi-messenger sources for sub-threshold kilonova and GW joint observations is 5 or 13 per year in this simulation. Considering the GW detection volume, which is inversely proportional to $\rho_{\star}^2$, sub-threshold GW observations can increase the probe volume by up to 9-17\%. In reality, a large fraction of detections will be by two-detectors only (LIGO-L/V), therefore their localization area can be notably larger than what we simulated here. Therefore, in actual EMC follow-up observations of GWs, especially for sub-threshold GWs, there will likely be more contamination and misidentifications than discussed in this work.
Consequently, at the same S/N threshold, the purity of multi-messenger sources will decrease. Thus, the suggested lowest threshold value can serve as a lower limit for the threshold.

For sGRB and GW joint observations, the joint detection rate is 0-0.25 yr$^{-1}$ and 0.5 yr$^{-1}$ for \textit{Fermi} GBM and GECAM in our simulation, respectively. Based on theoretical model, sGRBs are expected to be the first EM signals observed following the merger of BNS. Therefore, timely follow-up observations of sGRBs hold significant importance for observations in other wavebands. Based on the findings of this study, we recommend that the S/N threshold for GWs be lowered to 8.1-8.5 (corresponding to a FAR of 641-13900 per year) for sGRB and GW joint observations.

We do not take into account the contamination of the EMCs, which is also an important factor to consider in EMC follow-up \citep[][]{2018ApJ...858...18C}. Addressing misidentification or contamination issues in multi-messenger GW observations is extremely challenging yet crucial for the electromagnetic follow-up of sub-threshold GWs, as it significantly affects the relevance of sub-threshold GW observations. The simulations in this work were based on very simplistic assumptions and failed to account for numerous contamination scenarios. Nevertheless, these assumptions led us to conclude that the advantages of lowering the S/N threshold are currently minimal, as misidentifications compromise the reliability of many sub-threshold multi-messenger signals.

Conducting follow-up observations for each GW source can be resource-intensive. It is crucial to distinguish astrophysical GW sources from background noise. \citep[][]{2022CQGra..39q5004A} offers insights, indicating that astrophysical system populations generally feature nearly equal masses and moderate spins. However, the time required for parameter estimation may lead to missed opportunities in detecting the sGRBs associated with these GWs.

\section*{Acknowledgements}

This work is supported by the National Key R\&D Program of China (2021YFA0718500). SXY acknowledges the support by the Institute of High Energy Physics (Grant No. E25155U1).

\section*{Data Availability}

All the data generated via simulation and used in the analysis are available upon request. The code used to generate these data are open sources, \href{https://gw-universe.org/}{https://gw-universe.org/}.


\input{example.bbl}
\bibliographystyle{mnras}
\bibliography{example} 








\bsp	
\label{lastpage}
\end{document}